\RequirePackage{amsmath}
\documentclass{article}
\usepackage[T1]{fontenc}
\usepackage[utf8]{inputenc}
\usepackage[american]{babel}
\usepackage{calc}

\usepackage{color,graphicx,xcolor,svgcolor}
\usepackage{capt-of}
\usepackage[all]{xy}

\usepackage{float}
\makeatletter
\renewcommand\floatc@plain[2]{\setbox\@tempboxa\hbox{{\bfseries #1.} #2}
  \ifdim\wd\@tempboxa>\hsize {\@fs@cfont #1:} #2\par
    \else\hbox to\hsize{\hfil\box\@tempboxa\hfil}\fi}
\makeatother
\floatstyle{plain}
\newfloat{protocol}{htbp}{lop}
\floatname{protocol}{Protocol}

\usepackage{tabularx} 
\usepackage{arydshln} 
\usepackage{mdwlist}
\usepackage{multicol}
\usepackage{multirow}
\usepackage{rotating}

\usepackage{tikz,makecell}
\usetikzlibrary{arrows}
\usetikzlibrary{positioning}

\usepackage{xspace,amsfonts}
\newcommand{\mat}[1]{\ensuremath{\mathbf{#1}}}
\newcommand{\bigO}[1]{\ensuremath{O(#1)}\xspace}
\newcommand{\Z}{\ensuremath{\mathbb{Z}}\xspace}
\newcommand{\Q}{\ensuremath{\mathbb{Q}}\xspace}
\newcommand{\B}{\ensuremath{\mathbb{B}}\xspace}
\newcommand{\F}{\ensuremath{\mathbb{F}}\xspace}
\newcommand{\ZZ}{\Z}
\newcommand{\R}{\ensuremath{\mathbb{R}}\xspace}
\newcommand{\sdp}{semidefinite program\xspace}

\newcommand{\psd}{positive semidefinite\xspace}

\newcommand{\psdness}{positive semidefiniteness\xspace}

\newcommand{\symm}[2]{\mathbb{S}#1^{#2\times #2}}

\newcommand{\sos}{sum-of-squares\xspace}

\newcommand{\sample}[1]{\ensuremath{\xleftarrow{\$} {#1}}\xspace}
\newcommand{\checks}[1]{\ensuremath{\mathrel{\stackrel{?}{#1}}}}

\newcommand{\overlongrightarrow}[2]
{\vbox{\offinterlineskip
      \halign{##\cr
              \hfill #1 \hfill\cr
              \hbox to #2{\relax\rightarrowfill}\cr}}\ignorespaces
}

\newcommand{\overlongleftarrow}[2]
{\vbox{\offinterlineskip
      \halign{##\cr
              \hfill #1 \hfill\cr
              \hbox to #2{\relax\leftarrowfill}\cr}}\ignorespaces
}

\title{Proof-of-work certificates that can be efficiently
computed in the cloud}
\author{Jean-Guillaume Dumas\footnote{
  {Universit\'e Grenoble Alpes}.
  {Laboratoire Jean Kuntzmann}, CNRS, UMR 5224.
  {700~avenue centrale}, IMAG - CS 40700,
  {38058 Grenoble, cedex 9}
  {France}.
\href{mailto:Jean-Guillaume.Dumas@univ-grenoble-alpes.fr}{Jean-Guillaume.Dumas@univ-grenoble-alpes.fr}}}

\usepackage{hyperref}

\newcommand{\doi}[1]{\url{https://doi.org/#1}}
\definecolor{darkred}{rgb}{0.545098,0.,0.}
\definecolor{darkgreen}{rgb}{0,0.5,0}
\definecolor{darkblue}{rgb}{0.,0.,0.545098}
 \makeatletter
 \hypersetup{
 pdftitle={\@title},
 pdfauthor={Jean-Guillaume Dumas},
 breaklinks=true, 
 colorlinks=true,
 urlcolor=blue,
 citecolor=darkred,
 linkcolor=darkgreen,
 }
 \makeatother

\begin{document}
\maketitle
\begin{abstract}
In an emerging computing paradigm, computational capabilities, from
processing power to storage capacities, are offered to users over
communication networks as a cloud-based service.  
There, demanding computations are outsourced in order to limit
infrastructure costs.  

The idea of verifiable computing is to associate a data structure, a
\emph{proof-of-work certificate}, to the result of the outsourced
computation. This allows a verification algorithm to prove the validity
of the result, faster than by recomputing it. 
We talk about a Prover (the server performing the computations) and a
Verifier.  

Goldwasser, Kalai and Rothblum gave in 2008 a generic method to verify
any parallelizable computation, in almost linear time in the size of
the, potentially structured, inputs and the result. 
However, the extra cost of the computations for the Prover (and
therefore the extra cost to the customer), although only almost a
constant factor of the overall work, is nonetheless prohibitive in
practice.

Differently, we will here present problem-specific procedures in
computer algebra, e.g. for exact linear algebra computations,
that are Prover-optimal, that is that have much less financial
overhead.

\end{abstract}

\section{Introduction}
In an emerging computing paradigm, computational capabilities, from
processing power to storage capacities, are offered to users over
communication networks as a service. 

\begin{figure}[htbp]\centering
\includegraphics[width=\textwidth]{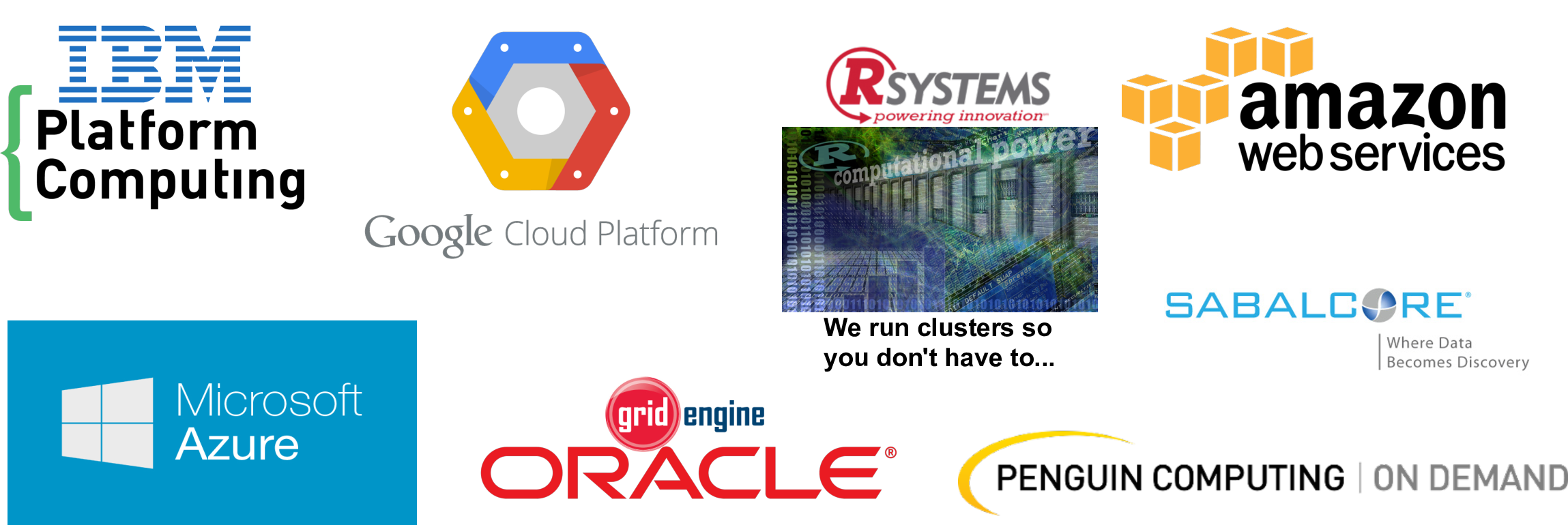}
\caption{Some outsourced computing services}\label{fig:CS}
\end{figure}

Many such outsourcing platforms are now well established, as Amazon web services
(through the Elastic Compute Cloud), Microsoft Azure, IBM Platform Computing or
Google cloud platform (via Google Compute Engine), as shown in
Figure~\ref{fig:CS}. 
None of these platforms,
however, offer any guarantee whatsoever on the calculation: no
guarantee that the result is correct, nor even that the computation
has even effectively been done.

\subsection{Verifiable computing}
This new paradigm holds enormous promise for increasing the utility of
computationally weak devices. A natural approach is for weak devices
to delegate expensive tasks, such as storing a large file or running a
complex computation, to more powerful entities (say servers) connected
to the same network. While the delegation approach seems promising, it
raises an immediate concern: when and how can a weak device verify
that a computational task was completed correctly? This practically
motivated question touches on foundational questions in cryptography,
coding theory, complexity theory, proofs and algorithms.
\begin{figure}[htbp]
\includegraphics[width=\textwidth]{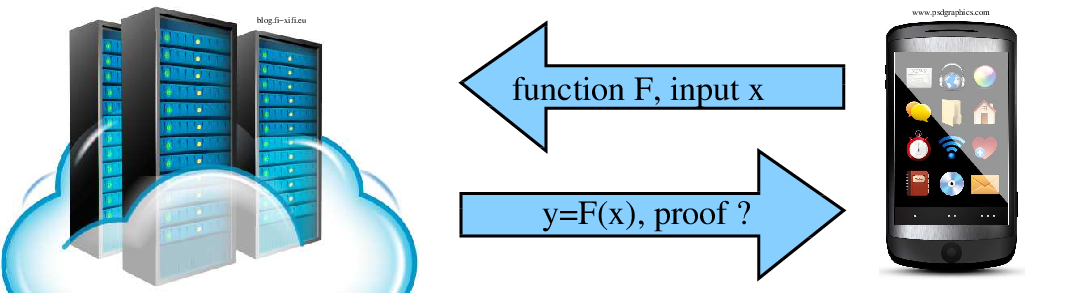}
\caption{Verifying the computation should take less time than
  computing it}\label{fig:deleg}
\end{figure}

More generally, the question of verifying a result at a lower cost
(time, memory) than that of recomputing it, as shown on
Figure~\ref{fig:deleg}, is of paramount importance. Another example of
application is for the extension of the trust about results computed via
probabilistic or approximate algorithms. There the idea is to gain confidence
into the correctness, but only at a cost negligible when compared to that of the
computation.

\subsection{Linear algebra, global optimization}\label{sec:artin}
For instance, GL7d19 is an $1\,911\,130 \times{} 1\,955\,309$ matrix 
whose rank $1\,033\,568$ was computed once in 2007 with a Monte-Carlo randomized
algorithm~\cite{jgd:2007:pasco}. This required $1050$ CPU days of computation. 
We thus need publicly verifiable certificates to improve our confidence in
computational results.

In linear algebra our original motivation is also related to \sos.
Indeed, by Artin's solution to Hilbert 17th Problem, any polynomial inequality
$\forall \xi_1,\ldots,\xi_n\in\R, f(\xi_1,\ldots,\xi_n)\geq
g(\xi_1,\ldots,\xi_n)$ can be proved by a fraction of \sos:
\begin{equation}\label{eq:sos}
  \exists u_i, v_j \in \R[x_1,\ldots,x_n],
  f-g=\left(\sum_{i=1}^\ell u_i^2\right)/\left(\sum_{j=1}^m v_j^2\right)
\end{equation}
Such proofs can be used to establish global minimality for

$g = \inf_{\xi_v\in\R} f(\xi_1$, $\ldots$, $\xi_n)$
and constitute certificates in non-linear global optimization.
A symmetric integer matrix $W\in \symm{\ZZ}{n}$ is \psd,
denoted by $W \succeq 0$, if all its eigenvalues, which then must be
real numbers, are non-negative. Then, a certificate for \psdness of rational
matrices constitutes, by its Cholesky factorizability, the final computer
algebra step in an exact rational \sos proof, namely
\begin{multline}\label{eq:sdp}
  \hspace*{-0.5em}\exists e\ge 0,\ W^{[1]}\succeq 0,\ W^{[2]} \succeq 0,\ W^{[2]}\neq \mathbf{0}:
  \\
  (f-g)(x_1,\ldots,x_n)\cdot(m_e(x_1,\ldots,x_n)^T W^{[2]}m_e(x_1,\ldots,x_n))=\\
  m_d(x_1,\ldots,x_n)^T W^{[1]}m_d(x_1,\ldots,x_n),
\end{multline}
where the entries in the vectors $m_d,m_e$ are the terms occurring in $u_i,v_j$
in (\ref{eq:sos}). 
In fact, (\ref{eq:sdp}) is the \sdp that one
solves~\cite{Kaltofen:2011:quadcert}. 
Then, the client can verify the positiveness by checking Descartes' rule of sign
on the {\em certified} characteristic polynomial of $W^{[1]}$ and $W^{[2]}$.
Thus arose the question how to give possibly probabilistically checkable
certificates for linear algebra problems.

\subsection{Techniques}
The tools used to provide such efficient \emph{proof-of-work certificates} stem
from programs that check their work~\cite{Blum:1995:checkwork}, to proof of
knowledge protocols~\cite{Bangerter:2005:pokdl}, via error-correcting
codes~\cite{Kaltofen:2014:interp,Gasieniec:2016:freivalds} complexity
theory~\cite{Aaronson:2009:algebrization} or secure multiparty
protocols~\cite{Cramer:2001:smc}, and the interaction of these different
methodologies is crucial.\\

Here we will thus follow this road-map:
\begin{itemize}
\item We recalled that global optimization can be reduced to linear
  algebra. Thereupon, we will focus on certificates for linear algebra
  problems~\cite{Kaltofen:2011:quadcert} in computer algebra. Those
  extend in particular the
  randomized algorithms of Freivalds~\cite{Freivalds:1979:certif}.
\item We combine those with probabilistic interactive proofs of
  Babai~\cite{Babai:1985:arthurmerlin} and Goldwasser et
  al.~\cite{Goldwasser:1985:IPclass}, 
\item as well as Fiat-Shamir
  heuristic~\cite{Fiat:1986:Shamir,Bellare:1993:randomoracle} turning
  interactive certificates into non-interactive heuristics subject to
  computational hardness.
\item Overall, we obtain problem-specific efficient
  certificates for dense, sparse, structured matrices with coefficients in
  fields or integral domains. 
\end{itemize}

\section{Interactive protocols, the PCP theorem and homomorphic encryption}

\subsection{Arthur-Merlin interactive proof systems}
A proof system usually has two parts, a theorem $T$ and a proof $\Pi$, and the
validity of the proof can be checked by a verifier~$V$.
Now, an {\em interactive proof}, or a
{\em $\sum$-protocol}, is a dialogue between a prover $P$ (or {\em Peggy} in
the following) and a verifier $V$ (or {\em Victor} in the following), where $V$
can ask a series of questions, or challenges, $q_1$, $q_2$, $\ldots$ and $P$ can
respond alternatively, in successive \emph{rounds}, with a series of strings
$\pi_1$, $\pi_2$, $\ldots$, the responses,  in order to prove the theorem $T$. 
The theorem is sometimes decomposed into two parts, the hypothesis, or input,
$H$, and the commitment,~$C$. Then the verifier can accept or reject the
proof: 
$V (H,C, q_1, \pi_1,q_2,\pi_2,\ldots)\in\{\text{accept}, \text{reject}\}$. 

To be useful, such proof systems should satisfy {\bf completeness} (the prover
can convince the verifier that a true statement is indeed true) and {\bf
  soundness}
(the prover cannot convince the verifier that a false statement is true). 
More precisely, the protocol is {\em complete} if the probability that a true
statement is rejected by the verifier can be made arbitrarily small.
Similarly, the protocol is {\em sound} if the probability that a false statement
is accepted by the verifier can be made arbitrarily small.
The completeness (resp. soundness) is {\em perfect} if accepted (resp. rejected)
statement are always true (resp. false). 

It turns out that interactive proofs with perfect completeness are as powerful
as interactive proofs \cite{Furer:1989:perfectcomp}.
Thus in the following, as we want to prove correctness of a result more than
proving knowledge of it, we will mainly show interactive proofs with perfect
completeness.

The class of problems solvable by an interactive proof system (IP) is equal to
the class PSPACE~\cite{Shamir:1992:IPPSPACE} and a probabilistically checkable
proof, PCP$\left[r(n),\pi(n)\right]{}$, for an input of length $n$, is a type
of proof that can be checked by a randomized algorithm using a bounded amount of
randomness $r(n)$ and reading a bounded number of bits of the proof $\pi(n)$. 
For instance,
PCP$\left[\bigO{\log{n}},\bigO{1}\right]$=NP~\cite{Babai:1990:pcp,Arora:1992:pcp}.

In general, interactive protocols encompass many kinds of proofs and Prover
and Verifier settings. One can think of the difficulty of integer factorization 
versus that of re-multiplying found factors.
Another example could be satisfiability checking, where the solver has to
explore the state space, while verifying a variable assignment or a conflict
clause could be much simpler~\cite{Abraham:2016:scsquare}. 
In computer algebra, the Prover can be a probabilistic algorithm or a
symbolic-numeric program, where the Verifier would perform the checks exactly or
symbolically; further, computer algebra systems could perform a complex
calculations where an interactive theorem prover (or proof assistant like
Isabel-HOL or Coq) only has to a posteriori formally verify the
certificate~\cite{Chyzak:2014:itp,Calude:2016:casc}. 

Table~\ref{tab:proververifier} gives more examples of such settings.
\begin{table}[htbp]\centering
\begin{tabular}{|l|l|}
\hline
Prover & Verifier\\
\hline
Computer Scientist & Mathematician \\
Computer Algebra system & Formal proof assistant\\
Cloud & User\\
Server & Client\\
Cellphone & Trusted platform module\\
\hline
\end{tabular}
\caption{Examples of Prover/Verifier settings}\label{tab:proververifier}
\end{table}

\subsection{Goldwasser et al. prover efficient interactive certificates}
Now, efficient protocols (interactive proofs between a
\emph{Prover}, responsible for the computation, and a \emph{Verifier}, to be
convinced) can be designed for delegating computational tasks. 

Recently, generic
protocols, mixing zero-sum checks~\cite{Lund:1992:zerosumcheck} and
probabilistically checkable proofs, have been 
designed by teams around 
Shafi Goldwasser at the MIT or Harvard, for circuits with
polylogarithmic depth~\cite{Goldwasser:2008:delegating,Thaler:2013:crypto},
namely for problems that can be efficiently solved on a parallel computer (in
the NC or AC complexity class).  
These results have also been extended to any structured
inputs (any polynomial-time-uniform polylog-depth Boolean circuits in the
sense of Beame's et al, \cite{BCH86}, division
circuits)~\cite{jgd:2017:polylog}. 

The resulting protocols are interactive and there is a
trade-off between the number of interactive rounds, the volume of communication
and the computational cost~\cite{Reingold:2016:stoc}; the cost for the verifier
being usually only roughly proportional to the input size.

These protocols can, e.g., certify that two supersparse polynomials are
relatively prime in verifier cost which is polylog time (and rounds) in the
degree.

The produced certificates, in analogy to processor-efficient parallel
algorithms, are Prover-efficient: if the cost to compute the result by the best
known algorithm is $T(n)$ for a size $n$ problem, then the cost to produce the
result together with the verifiable certificate is $T(n)^{1+o(1)}$. 

Those techniques can however produce a non negligible practical overhead for the
Prover and are restricted to certain classes of circuits. 

\subsection{Parno et al. homomorphic solutions}

Another approach as been developed by Gentry et
al., at Microsoft and IBM research, it is
\href{http://research.microsoft.com/en-us/projects/verifcomp}{Pinocchio}.
It solves a broader range of problems, to the cost of using relatively
inefficient homomorphic
routines~\cite{Parno:2013:Pinocchio} in an amortized
way. 

The idea is that the Prover should run the program (or at least part of the
program twice), once normally on the input, and once homomorphically on an
encrypted version of the input. The Verifier will then verify the consistency
between the normal output and the encrypted one. Usually the Verifier is
required to run the algorithm at least once for a given size or structure of the
input but can reuse this for multiple inputs.

This generic procedure can be applied on specific linear algebra or polynomial
problems~\cite{Fiore:2012:PVD,Blanton:2014:matrix,Elkhiyaoui:2016:ETPV,jgd:2017:acisp},
or on generic quadratic arithmetic programs~\cite{Parno:2013:Pinocchio}.
There, fully homomorphic encryption can be used~\cite{Gentry:2014:nizkfhe} or
sometimes just pairings~\cite{Parno:2013:Pinocchio} and/or cryptographic
hashes~\cite{Fiore:2016:hashandprove}. 

Here also the Prover can be efficient, but subject in practice to the overhead
of homomorphic computations.

\subsection{Public verification, delegatability and zero-knowledge}
Interactive certificates require some exchanges between the Prover and the
Verifier. With such a protocol, the Verifier can be \emph{privately} convinced
that the computation of the Prover produced the correct answer. 
This does not mean that other people would be convinced be the transcript of
their exchange: the Prover and Verifier could be in cahoots and the supposedly
random challenges carefully crafted. 

Fiat-Shamir heuristic~\cite{Fiat:1986:Shamir,Bellare:1993:randomoracle} can thus
turn interactive certificates into non-interactive heuristics subject to
computational hardness: the random challenges are replaced by cryptographic
hashes of all previous data and exchanges. Crafting such values would then
reduce to being able to forge cryptographic
fingerprints~\cite[\S~4.5]{jgd:2014:interactivecert}.

Further, more properties could be sought for such protocols, such \emph{privacy}
of data and/or computations.
In this setting, a publicly verifiable computation scheme can also be four
algorithms (\emph{KeyGen}, \emph{ProbGen}, \emph{Compute}, \emph{Verify}), where
\emph{KeyGen} is some (amortized) preparation of the data, \emph{ProbGen} is the
preparation of the input, \emph{Compute} is the work of the \emph{Prover} and
\emph{Verify} is the work of the \emph{Verifier}~\cite{Parno:2012:delegate}. 
Usually the Verifier also executes \emph{KeyGen} and \emph{ProbGen} but in a
more general setting these can be performed by different entities (respectively
called a \emph{Preparator} and a \emph{Trustee}). 

This allows to define several adversary models but usually the protocols are
secure against a \emph{malicious Prover only} (that is the Client must trust
both the Preparator and the Trustee). 

One can also further impose that there is no interaction between the Client and
the Trustee after the Client has sent his input to the Server. Publicly
verifiable protocols with this property are said to be \emph{publicly
  delegatable}~\cite{Blanton:2014:matrix,Elkhiyaoui:2016:ETPV,jgd:2017:acisp}.

Then, some different properties of the protocol could be desirable, such as not
disclosing the result but instead just providing a \emph{proof-of-work}. This
results in general in \emph{zero-knowledge} protocols over confidential data,
such as cryptocurrency transactions, as
in, e.g.,~\cite{Goldwasser:1985:IPclass}, with recent efficient
implementations~\cite{Bootle:2016:eurocrypt,Ben-Sasson:2017:BCGGH,cryptoeprint:2018:046,Bunz:2018:bulletproof}.

\subsection{Problem-specific efficient certificates}
Differently, dedicated certificates (data structures and
algorithms that are verifiable a posteriori, without interaction) have
also been developed, e.g., in computer algebra for exact linear
algebra~\cite{Freivalds:1979:certif,Kaltofen:2011:quadcert,jgd:2014:interactivecert,jgd:2016:gammadet,jgd:2017:rpmcert},
even for problems that are not structured. 
There the certificate constitute a proof of correctness of a result,
not of a computation, and can thus also \textbf{detect bugs} in the
implementations. 

\begin{figure}[htbp]\center
\fbox{\includegraphics[width=\textwidth/2]{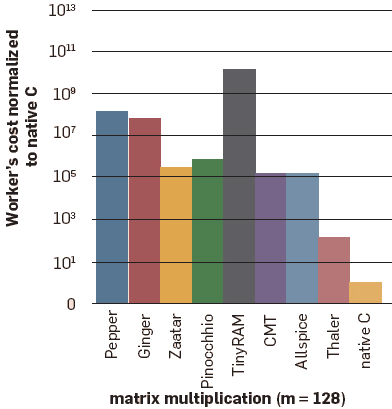}}\bigskip

\begin{tabular}{|l||r|r|}
\hline
$2048\times{}2048$ & Thaler\cite{Thaler:2013:crypto} & Ad-hoc
~\cite{Freivalds:1979:certif} \\ 
\hline
Server time & 18.23s & 0.65s\\
Certificate overhead & 0.13s & 0.00s\\
Client time & 2.89s & 0.01s\\
\hline
\end{tabular}
\caption{Generic protocols~\cite{Walfish:2015:VCWRT} versus dedicated protocols for matrix multiplication}\label{tab:matmul}
\end{figure}

Moreover, problem-specific certificates can gain crucial logarithmic factors for
the verifier and allow for optimal prover computational time, see
Figure~\ref{tab:matmul}.  

For this, the main difficulty is to be able to design verification
algorithms for a problem that are completely orthogonal to the computational
algorithms solving it, while remaining checkable in time and space not
much larger than the input.

\section{Prover-optimal certificates in linear algebra}
We show in this section, that such problem-specific certificates are attainable
in linear algebra, where we allow certificates that are validated by Monte Carlo
randomized algorithms.
\subsection{Freivalds zero equivalence of matrix
  expressions}\label{ssec:freivalds}
The seminal certificate in linear algebra is due to R{\=u}si{\c{n}}{\v{s}}
Freivalds~\cite{Freivalds:1979:certif}:
quadratic time is feasible because a matrix multiplication $AB$ can be certified
by the resulting product matrix $C$ via the probabilistic projection to matrix-vector products
(see also \cite{Kimbrel:1993:PAV} who reduced the requirements to only
$\bigO{\log(n)}$ random bits), shown in Protocol~\ref{fig:freivalds}.

\begin{protocol}[htbp]
\centerline{
\framebox{
    \begin{tabular}{@{}r@{}l@{}@{}c@{}c@{}cl@{}}
& \multicolumn{1}{c}{{\itshape Prover}}
        && \hbox to 0pt {\hss \itshape Communication \hss} &&
\multicolumn{1}{c}{{\itshape Verifier}}
\\ \hline
\\
 & && $\mat{A}\in\F^{m{\times}k}$, $\mat{B}\in\F^{k{\times}n}$ &\\
\\ \hline
\rule{0pt}{4ex}
  &Compute $\mat{C}=\mat{A}\cdot{}\mat{B}$&&\overlongrightarrow{$\mat{C}$}{5em}&& $r\sample{S}\subseteq\F$\\
  &&&&& Form $\vec{v}=[1,r,r^2,\ldots,r^{n-1}]^T$\\
  &&&&& $\mat{A}\left(\mat{B}\vec{v}\right)-\mat{C}\vec{v}\checks{=} \vec{0}$\\
\end{tabular}
}
}
\caption{Matrix multiplication certificate~\cite{Kimbrel:1993:PAV}.}\label{fig:freivalds}
\end{protocol}

In Protocol~\ref{fig:freivalds}, we give the variant of~\cite{Kimbrel:1993:PAV}
that requires $\log(n)$ random bits, but works over sufficiently large
coefficient domains, as its soundness is $1-\frac{|S|}{n}$ by the
DeMillo-Lipton/\allowbreak Schwartz/\allowbreak Zippel
lemma~\cite{DeMillo:1978:ipl,Zippel:1979:ZSlemma,Schwartz:1979:SZlemma}. 
Freivalds original version randomly selects a zero-one vector instead. This
requires $n$ random bits instead but applies to any ring and gives a soundness
larger than $\frac{1}{2}$. 

In both cases it is sufficient to repeat the test several times to achieve any
desired probability.

\subsection{Reductions to matrix multiplication}\label{ssec:kns}
With a certificate for matrix expressions, then one can certify {\bf any
algorithm that reduces to matrix multiplication}: the Prover records all the
intermediate matrix products and sends them to the Verifier who reruns the same
algorithm but checks the matrix products instead of computing
them~\cite{Kaltofen:2011:quadcert}, as shown in Protocol~\ref{fig:kns}.
\begin{protocol}[htbp]
\centerline{
\framebox{
    \begin{tabular}{@{}r@{}l@{\hspace{10pt}}@{}c@{}c@{\hspace{10pt}}cl@{}}
& \multicolumn{1}{c}{{\itshape Prover}}
        && \hbox to 0pt {\hss \itshape Communication \hss} &&
\multicolumn{1}{c}{{\itshape Verifier}}
\\ \hline
\rule{0pt}{4ex}
  &&& All intermediate &&Runs the algorithm but\\
  &Runs the algorithm && matrix products && replace each matrix products\\
  &&&$\xrightarrow{\hspace{70pt}}$&& by Freivalds' checks\\
\end{tabular}
}
}
\caption{Certificates with reduction to matrix
  multiplication~\cite[\S~5]{Kaltofen:2011:quadcert}.}\label{fig:kns}
\end{protocol}

Overall, the communications and Verifier computational cost are given by taking
$\omega=2$ in the Prover's complexity bounds (with potential additional
logarithmic factors due to summations).
Further, {\emph the production of the certificate has no computational overhead}
for the Prover: it only adds the communication of the intermediate matrix
products. 

For instance, Storjohann's Las Vegas rank algorithm of integer
matrices~\cite{Storjohann:2009:IMR} becomes a non-interactive/non-cryptographic
Monte Carlo checkable proof-of-work certificate that has soft-linear time
communication and verifier bit complexity in the number of input bits!

\subsection{Sparse or structured matrices}\label{ssec:sparse}
When the matrices are sparse or present some structure, quadratic run time
and/or quadratic communications might be overkill for the Verifier. 
There it is better if his communications and computational cost is of the form
$\mu(A)+n^{1+o(1)}$ where $\mu(A)$ is the number of operations to perform a
matrix-vector products. This scheme is thus also interesting if the considered
matrix is only given as a blackbox~\cite{KaTr90}.\\

In that vein, we now have certificates for :
\begin{itemize}
\item \textbf{non-singularity}, Protocol~\ref{fig:nonsing};

\begin{protocol}[!p]\center
  \noindent\resizebox{\linewidth}{!}{$$
    \xy
    \xymatrix@R=7pt@C=1pc@W=15pt{
      &Prover &&&& Verifier\\
      Input&&& \mat{A}\in\F^{n\times n} &&\\
      \ar@{.}[rrrrrr]&&&&&&\\
      Commitment& & \ar[rr]^*[*1.]{1:\ \text{non-singular}} &&&\\
      Challenge& & && \ar[ll]_*[*1.]{2:\ \vec{b}}& \vec{b}\sample{S^n}\subset\F^n\\
      Response& \vec{w}\in\F^n&\ar[rr]^*[*1.]{3:\ \vec{w}}&&& \mat{A}\vec{w}\checks{=}\vec{b} \\
    }\POS*\frm{-}
    \endxy
    $$
  }
  \caption{Blackbox interactive certificate of non-singularity~\cite{jgd:2014:interactivecert}}\label{fig:nonsing}
\end{protocol}

\item \textbf{an upper bound to the rank}, Protocol~\ref{fig:upperbound} (if elimination
  on the input matrix is possible for the Prover then a variant without
  preconditioners can be used~\cite{Eberly:2015:rankcert,jgd:2017:rpmcert});  

\begin{protocol}[!p]
\center
  \noindent\resizebox{\linewidth}{!}{$$
    \xy
    \xymatrix@R=4pt@C=0pc@W=10pt{
      &Prover &&&& Verifier\\
      &&& A\in\F^{m\times n} &&S\subset\F\\
      \ar@{.}[rrrrrr]&&&&&&\\
      &rank(A) \leq r &\ar[rr]^*[*1.]{1:\ r} &&  & r \checks{<} \min\{m,n\}\\
      & & & & \ar[ll]_*[*1.]{2:\ U,V}&U\in\B_S^{m\times m},V\in\B_S^{n\times
        n}\\
      & & & & & \text{preconditioners of size}~n^{1+o(1)}\\
      & 
      w\in\F^{r+1}\neq 0&\ar[rr]^*[*1.]{3:\ w}&&& 
      w \checks{\neq}0\\
      &&&&&\left[ I_{r+1}|0 \right] U A V \ensuremath{\left[\begin{array}{cc}I_{r+1}\\0\end{array} \right]} w \checks{=}0
      \\
    }\POS*\frm{-}
    \endxy
    $$
  }
  \caption[Blackbox interactive certificate for an upper bound to the
    rank]{Blackbox upper bound to the rank certificate~\cite{jgd:2014:interactivecert}}\label{fig:upperbound}
\end{protocol}

\item the \textbf{rank}, combining Protocols~\ref{fig:nonsing} and~\ref{fig:upperbound};

\item the \textbf{minimal polynomial}, using Protocol~\ref{fig:fAuvcert} (where
  $f_u^{A,v}$ is the monic scalar minimal generating polynomial of the sequence 
  $u^T v, \ldots, u^T A^i v$, $\rho_u^{A,v}$ is such that
  $\rho_u^{A,v}=f_u^{A,v} {\cdot} G$ with $G$ the generating function of the
  latter sequence, for random vectors $u$ and $v$, chosen by the
  Verifier~\cite[Theorem~5]{Kaltofen:1995:ACB});  

\begin{protocol}[!htbp]
\centerline{%
\begin{tabular}{
|
@{\hspace*{0.2em}}r%
@{\hspace{0.35em}}l%
@{}@{}c%
@{\hspace*{0.6em}}@{}l%
@{\hspace*{0.2em}}
|
}
\hline
& \multicolumn{1}{c}{{\itshape Prover}}
& \hbox to 0em{\hss \itshape Communication \hss} &
\multicolumn{1}{c|}{{\itshape Verifier}}
\\ 
\hline
\rule{0pt}{10pt}%
&
$H(\lambda)=f_u^{A,v}(\lambda)$,& &\\ 	
&$h(\lambda)=\rho_u^{A,v}(\lambda)$.&  	
\overlongrightarrow{$H,h$}{30pt}	
&
\\[1ex]
&
$\phi, \psi \in \F[\lambda]$ with
&&
\\
&$\phi f_u^{A,v} + \psi \rho_u^{A,v} = 1,$
& \overlongrightarrow{$\phi,\psi$}{30pt}
& $\deg(\phi)\checks{\le}\deg(h) - 1$,
\\
& & &
$\deg(\psi) \checks{\le} \deg(H)- 1.$
\\
&
&
& Random $r_0 \in S\subseteq \F$.
\\
& & &  Checks  $\text{GCD}(H(\lambda), h(\lambda))= 1$ by $\phi(r_0) H(r_0)+ \psi(r_0) h(r_0)\checks{=} 1.$
\\
& Computes $w$ such that
& \overlongleftarrow{$r_1$}{30pt}
& Random $r_1 \in S\subseteq \F$.   
\\
&$(r_1 I_n-A)w = v$.
& \overlongrightarrow{$w$}{30pt}
& Checks
$(r_1 I_n-A)w \checks{=} v$ and $(u^T w) H(r_1)\checks{=}h(r_1)$.\\[1ex]	
&&& Returns $f_u^{A,v}(\lambda)=H(\lambda)$.	
\\[1ex] \hline
\end{tabular}
}
\caption{\label{fig:fAuvcert}
Certificate for $f_u^{A,v}$~\cite{jgd:2016:gammadet}}
\end{protocol}


\item the \textbf{determinant}, Protocol~\ref{fig:WiedDet}, which randomness
  could be reduced from $\bigO{n}$ to a constant number of field
  elements~\cite[\S~7]{jgd:2016:intdet}.

\begin{protocol}[!htbp]
\centerline{
\framebox{
    \begin{tabular}{@{}r@{\hspace{0.35em}}l@{\hspace{10pt}}@{}c@{}c@{}c@{}c@{}}
& \multicolumn{1}{c}{{\itshape Prover}}
        && \hbox to 0pt {\hss \itshape Communication \hss} &&
\multicolumn{1}{c}{{\itshape Verifier}}
\\ \hline
\rule{0pt}{4ex}
1.&Form $B=DA$ with&& &&\,\hspace{100pt}\,\\
  &$D\in S^{n}\subseteq \F^{*n}$&&\overlongrightarrow{$D,u,v$}{60pt} &&\\
  &and $u,v\in S^n$, &&\\
  &s.t. $\deg(f_u^{B,v})=n$. &&\\
\cline{3-3}
\cline{5-5}
&&\multicolumn{1}{|c}{\hskip 0.25em}&\smash{\raisebox{1ex}{Protocol~\ref{fig:fAuvcert}}}&
\multicolumn{1}{c|}{\hskip 0.25em}\\
2.&&\multicolumn{3}{|c|}{\overlongrightarrow{$H,h,\phi,\psi$}{60pt}}&{Checks:}\\
3.&&\multicolumn{3}{|c|}{\overlongleftarrow{$r_1$}{60pt}}&{$\deg(H)\checks{=}n$,}\\
4.&&\multicolumn{3}{|c|}{\overlongrightarrow{$w$}{60pt}}&{$H\checks{=}f_u^{B,v}$, w.h.p.}\\
\cline{3-5}
\\
5.&&&&& {Returns $\displaystyle\frac{f_u^{B,v}(0)}{\det(D)}$.}
\end{tabular}
}
}
\caption{\label{fig:WiedDet}
Determinant certificate for a
non-singular blackbox~\cite{jgd:2016:gammadet}}
\end{protocol}

\end{itemize}

Additionally, properties of the given matrices can also sometimes be discovered
at low cost: whether the blackbox is a \textbf{band matrix}, 
has a \textbf{low displacement rank}, 
has a few or many \textbf{nilpotent blocks} or \textbf{invariant
  factors}~\cite{Eberly:2016:certifs}. Similarly, the existence of a 
\textbf{triangular one sided equivalence}, as well as the \textbf{rank profiles}
can also be certified without sending an explicit factorization to the
Verifier~\cite{jgd:2017:rpmcert}. For the latter, the price to pay is to require
a linear number of rounds.

\subsection{Integer or polynomial matrices}\label{ssec:integral}

Over an integral domain, the verification procedure can be performed via a
randomly chosen modular projection. If there are sufficiently many \emph{small}
maximal ideals, then one can uniformly chose one at random and then ask for a
certification of the result in the associated quotient field as shown in
Protocol~\ref{fig:quotient}.

\begin{protocol}[htb]
\centerline{
\framebox{
    \begin{tabular}{c@{\hspace{10pt}}r@{}l@{\hspace{10pt}}@{}c@{}c@{\hspace{10pt}}cl@{}}
&& \multicolumn{1}{c}{{\itshape Prover}}
        && \hbox to 0pt {\hss \itshape Communication \hss} &&
\multicolumn{1}{c}{{\itshape Verifier}}
\\ \hline
\rule{0pt}{4ex}
\emph{Commitment} &&Result $r\in\R$ &&\overlongrightarrow{$r$}{50pt}&& \\
\emph{Challenge}   && &&\overlongleftarrow{$\mathcal I$}{50pt}&& ${\mathcal I}\sample{\text{maximal ideals}}$\\
\emph{Response}   &&Result $x\in\R/{\mathcal I}$ with &&
\overlongrightarrow{$x,{\mathcal C}_{\R/I}$}{50pt}&& $x\checks{\equiv} r\mod {\mathcal I}$ and\\
  &&field certificate ${\mathcal C}_{\R/I}$ && && ${\mathcal C}_{\R/I}(x)\checks{=}$ valid\\
\end{tabular}
}
}
\caption{Certification in a quotient field~\cite[\S~3.2 and \S~4.4]{jgd:2014:interactivecert}.}\label{fig:quotient}
\end{protocol}

For instance this gives very efficient certificates for polynomial or
integer/ra\-tio\-nal matrices, provided that one has a bound on the degree or the
magnitude of the coefficients:
\begin{itemize}
\item For \textbf{integral matrices}, if the true result $v$ is bounded in magnitude,
  then only a finite number of prime numbers will divide the difference between
  the commitment $r$ and the result. Therefore the result can be checked over a
  \emph{small} prime field~\cite[Theorem~5]{jgd:2014:interactivecert}.
\item For \textbf{polynomial matrices}, if the true $v(X)$ result's degree is
  bounded, then only a finite number of evaluation points can be roots of the
  difference polynomial between the committed one $r(X)$ and the result.  
  Therefore the result can be checked in the ground field at a \emph{small}
  evaluation point~\cite[Theorem~2]{jgd:2014:interactivecert}.
\end{itemize}

The latter results allows, for instance, to certify the global optimization
problems of Section~\ref{sec:artin}.

This is illustrated in Figure~\ref{fig:SoA}, where many of the reductions
presented  here are recalled.
\begin{figure}[p]\footnotesize
  \begin{sideways}
    \newlength{\aboveheight}
\setlength{\aboveheight }{30pt}
\begin{tikzpicture}[->,>=stealth',shorten >=1pt,auto,
  node distance=120pt,
  thick,
  align=center,
  dense node/.style={rectangle,rounded corners,fill=blue!20,draw,font=\sffamily},
  int node/.style={rectangle,rounded corners,fill=green!20,draw,font=\sffamily},
  bb node/.style={rectangle,rounded corners,fill=red!20,draw,font=\sffamily},
  gen node/.style={rectangle,rounded corners,fill=yellow!80,draw,font=\sffamily}]
  \node[dense node] (0) {{\sc MatVecMul}($\F$)};
  \node[dense node] (1) [above=\aboveheight of 0] {MM($\F$)};
  \node[dense node] (3) [above=\aboveheight of 1] {{\sc Inverse}($\F$)};
  \node[dense node] (2) [right of =3] {PLUQ($\F$)};
  \node[dense node] (4) [above=\aboveheight of 2] {{\sc Rank}($\F$)};
  \node[dense node] (11) [above=\aboveheight of 3] {{\sc NormalForm}($\F$)\\
					w. change of base};

  \node[gen node] (5) [right= of 4] {{\sc Det}};

  \path[every node/.style={font=\sffamily\small}]
    (5) edge node [left=7pt] {$Det(U)$} (2);

  \node[bb node]  (210) [below=\aboveheight of 5] {{\sc MinPoly}($\F$)};
  \node[bb node]  (215) [below=\aboveheight of 210] {{\sc KrylovSeq}($\F$)};
  \node[bb node]  (220) [right of =215] {{\sc LinSys}($\F$)};
  \node[bb node]  (225) [right of =5] {{\sc Rank}($\F$)};
  \coordinate[right= of 0] (r0);
  \coordinate[right= of r0] (r1);
  \node[bb node]  (217) [right=20pt of r1] {{\sc MatVecMul}($\F$)};

  \path[every node/.style={font=\sffamily\small}]
    (5) edge node [left=2pt,near end] {\cite{Wiedemann:1986:SSLE}} (210);

  \node[gen node] (51) [above left=40pt of 5] {{\sc CharPoly}};
  \node[int node] (10) [above=\aboveheight of 51] {{\sc CharPoly}($\Q$)};
  \node[int node] (7) [above=\aboveheight of 10] {{\sc Signature}($\Q$)};
  \node[int node] (8) [left of =7] {{\sc Sum-of-Squares}($\Q$)};
  \node[int node] (9) [above=\aboveheight of 8] {{\sc Global Optimization}($\Q$)};
  \node[int node] (c11) [above=\aboveheight of 11] {{\sc NormalForm}($\Q$)};
  \node[int node] (c4) [above=\aboveheight of 4] {{\sc Rank}($\Q$)};

  \node[int node] (c15) [above=\aboveheight of 5] {{\sc Det}($\Q$)};

  \path[every node/.style={font=\sffamily\small}]
    (8) edge node [above] {\cite{Kaltofen:2011:quadcert}} (7)
    (7) edge node [right] {[Descartes 1627]} (10)
    (8) edge node [right] {} (7)
    (7) edge node [right] {} (10)
    (1) edge node [left] {\cite{Freivalds:1979:certif}} (0)
    (2) edge [bend left=20] node [above left] {{\color{black}{$L \cdot U \stackrel{?}{==} A$}}} (1)
    (2) edge [bend left=40] node [right=5pt] {\cite{jgd:2017:rpmcert}} (0)
    (4) edge node [left] {$Rank(U)$} (2)
    (3) edge node [left=5pt] {{\color{black}{$A^{-1}\cdot A
          \stackrel{?}{==} I$}}} (1)

    (51) edge node [left=5pt,near end] {\cite{jgd:2014:interactivecert}} (5)
    (11) edge node [left] {\cite{Kaltofen:2011:quadcert}} (3)
    (7) edge node [left] {} (10)
    (8) edge node [left] {}  (7)
    (9) edge node [right] {[Artin'27]}  (8)
    (210) edge node [left] {\cite{Wiedemann:1986:SSLE}} (215)
    (215) edge node [left=5pt] {\cite{jgd:2016:intdet}} (217)
    (5) edge [bend left=20, near end] node [left=3pt,near end] {\cite{jgd:2016:gammadet}} (220)
    (225) edge node [right] {\cite{jgd:2014:interactivecert}} (220)
    (220) edge node [right] {{\color{black}{$Ax \stackrel{?}{==} b$}}} (217)

    (10) edge [dashed] node [right] {\cite{jgd:2014:interactivecert}} (51)
    (c11) edge [dashed] node [left] {} (11)
    (c4) edge [dashed] node [right] {} (4)

    (c15) edge [dashed] node [right] {} (5)

;

\end{tikzpicture}
  \end{sideways}
\caption{Global optimization via problem-specific interactive certificates:
  dense (purple) or sparse (red) algebraic problems, as well as over the reals
  (green) or oblivious (yellow).}\label{fig:SoA}  
\end{figure}

\subsection{Non-interactive certificates}
The certificates in Sections~\ref{ssec:freivalds} and~\ref{ssec:kns} are
non-interactive: all the communications can be recorded and publicly verified later.

On the contrary the certificates of Sections~\ref{ssec:sparse},
\ref{ssec:integral} are interactive: the Verifier chooses some random bits
during the computation of the certificate. Non-interactivity can be recovered
via Fiat-Shamir scheme: any random bits are generated by cryptographic hashes of
the inputs and all the previous intermediate commitments. Soundness is then
subject to standard cryptographic assumptions. 

For sparse or structured problems fewer results exists without this
assumption, or with worse complexity bounds:
\begin{itemize}
\item For the minimal polynomial (scalar or matrix) or the determinant,
  non-interactive certificates 
  exists, but with communications and computational cost \bigO{n\sqrt{\mu(A)}}
  instead of $\mu(A)+n^{1+o(1)}$~\cite{jgd:2016:intdet}.
\item Non-interactive certificates can also verify polynomial minimal approximant
  bases in \bigO{mD+m^{\omega}}, where $D$ is the sum of the column degrees of
  the output~\cite{Giorgi:2018:approximantcertif}.
\end{itemize}

\section{Some open problems}
We conclude this survey with some open problems in the area of problem specific
linear algebra certificates:

\begin{itemize}
\item {\bf Sparse Smith form}: for dense matrices, one can interactively certify
  any normal form via a Freivalds certificate on a randomly chosen modular
  factorization. With sparse matrices, even the modular projection of the change
  of base can be too large. In that setting extending protocols for the rank or
  the determinant to deal with the Smith form should be possible.

\item {\bf Non integral domains certificates}: more generally, how to
  efficiently certify some properties when there is no quotients or if those
  properties do not carry over (e.g., Smith form)?
\item We have defined certificates resisting a malicious server with unbounded
  power. This is error detection with unbounded number of errors. Thus the
  question of the complexity of {\bf problem specific unbounded error correction} also
  arises. This path again was first taken for matrix
  multiplication~\cite{Gasieniec:2016:freivalds} and was recently extended to
  the matrix inverse~\cite{Roche:2018:pagh}. 
\end{itemize}

\section*{Acknowledgment} 
I thank
Brice Boyer%
, Pascal Lafourcade%
, Shafi Goldwasser%
, Erich Kaltofen%
, Julio López Fenner%
, David Lucas%
, Vincent Neiger%
, Jean-Baptiste Orfila%
, Cl\'ement Pernet%
, Maxime Puys%
, Jean-Louis Roch%
, Dan Roche%
, Guy Rothblum%
, Justin Thaler%
, Emmanuel Thom\'e%
, Gilles Villard%
, Lihong Zhi
and an anonymous referee
for their helpful comments. 
\bibliographystyle{plainurl}
\bibliography{jgdbibl}

\end{document}